\documentclass[fleqn,usenatbib]{mnras}

\usepackage{newtxtext,newtxmath}

\usepackage[T1]{fontenc}
\usepackage{ae,aecompl}

\usepackage{graphicx}	
\usepackage{amsmath}	
\usepackage{xcolor}          

\title[Detection of cesium]{Detection of cesium in the atmosphere of the hot He-rich white dwarf HD~149499B}

\author[P. Chayer et al.]{
P. Chayer,$^{1}$\thanks{E-mail: chayer@stsci.edu}
C. Mendoza,$^{2}$
M. Mel\'endez,$^{1}$
J. Deprince$^{3,4}$
and J. Dupuis$^{5}$
\\
$^1$Space Telescope Science Institute, 3700 San Martin Drive, Baltimore, MD 21218, USA\\
$^{2}$Physics Center, Venezuelan Institute for Scientific Research (IVIC), Caracas 1020, Venezuela\\
$^{3}$Cahill Center for Astronomy and Astrophysics, California Institute of Technology, Pasadena, CA 91125, USA \\
$^{4}$Physique Atomique et Astrophysique, Universit\'{e} de Mons -- UMONS, 7000 Mons, Belgium\\
$^{5}$Canadian Space Agency, 6767 Route de l'A\'{e}roport, Saint-Hubert, QC J3Y 8Y9, Canada
}

\date{Accepted 2022 October 26. Received 2022 September 26; in original form ZZZ}

\pubyear{2022}

\begin{document}
\label{firstpage}
\pagerange{\pageref{firstpage}--\pageref{lastpage}}
\maketitle

\begin{abstract}
We report the first detection of cesium ($Z = 55$) in the atmosphere of a white dwarf. Around a dozen absorption lines of \ion{Cs}{iv}, \ion{Cs}{v}, and \ion{Cs}{vi} have been identified in the {\it Far Ultraviolet Spectroscopic Explorer} spectrum of the He-rich white dwarf HD~149499B ($T_{\rm{eff}} = 49$,500 K, $\log g = 7.97$). The lines have equivalent widths ranging from 2.3 to 26.9 m\AA.  We performed a spectral synthesis analysis to determine the cesium content in the atmosphere. Non-LTE atmosphere models were computed by considering cesium explicitly in the calculations.  For this purpose we calculated oscillator strengths for the bound--bound transitions of \ion{Cs}{iv}--\ion{Cs}{vi} with both \textsc{autostructure} (multiconfiguration Breit--Pauli) and {\sc grasp2k} (multiconfiguration Dirac--Fock) atomic structure codes as neither measured nor theoretical values are reported in the literature.  We determined a cesium abundance of $\log N{\rm{(Cs)}}/N{\rm{(He)}} = -5.45\pm0.35$, which can also be expressed in terms of the mass fraction $\log X_{\rm{Cs}} = -3.95\pm0.35$. 

\end{abstract}




\section{Introduction}

Because of their high gravity, white dwarfs are known to have atmospheres of either pure hydrogen or pure helium. Gravitational settling is so efficient in the atmospheres of these stars that element separation takes place rapidly in the absence of competing mixing mechanisms. The lightest element will float on the surface to form the atmosphere while the heavier will sink and disappear from sight. Although a small fraction of white dwarfs show traces of elements heavier than hydrogen and helium, the detection of trans-iron elements ($Z > 28$) in the spectra of the hot H-rich white dwarfs (DA) G191-B2B, Feige~24, and GD~246 by \citet{vennes_etal2005apj} and in the spectra of the hot He-rich white dwarfs (DO) HD~149499B and HZ~21 by \citet{chayer_etal2005apjl} was unexpected as these elements have extremely low solar abundances \citep{asplund_etal2009, grevesse_etal2015}.



In a series of papers spanning almost a decade, T. Rauch, K. Werner, and collaborators reported the detection of several other trans-iron elements in the spectra of G191-B2B and the DO white dwarf RE~0503$-$289 \citep{rauch_etal2012, rauch_etal2014_564,rauch_etal2014_566, rauch_etal2015_577_6, rauch_etal2015_577_88, rauch_etal2016_587, rauch_etal2016_590, rauch_etal2017_599, rauch_etal2017_606, rauch_etal2020_637, werner_etal2012_753, werner_etal2018_614,hoyer_etal2017}. They did not only identify these elements, but also calculated oscillator strengths for ionization stages \textsc{\lowercase{IV}}--\textsc{\lowercase{VII}} needed to determine the atmospheric abundances. This calculation was necessary to build reliable stellar atmospheric models since, in most cases, only a small number of oscillator strengths were available in the literature \citep[e.g.,][]{morton_2000}. By using these atomic data and a diffusion--radiative levitation code developed by \citet{dreizler_wolff1999aa}, \citet{rauch_etal2016_587} demonstrated that radiative levitation can hold up trans-iron elements in the atmospheres of G191-B2B and RE~0503$-$289. They also elucidated, at least qualitatively, that the abundances in RE~0503$-$289 are higher than those in G191-B2B because of a higher effective temperature. \citet{hoyer_etal2018} extended the search for trans-iron elements in the atmospheres of two other DOs and a PG1159 star, and suggested that the abundance diversity was due to the effects of diffusion and radiative levitation. More recently, \citet{lobling_etal2020} discovered trans-iron elements in the DAO white dwarf BD$-$22$^{\circ}$3467. It is interesting to note that all trans-iron elements observed in hot white dwarfs are overabundant. 

Eighteen trans-iron elements have been so far identified in the atmospheres of hot white dwarfs: Cu ($Z = 29)$, Zn (30), Ga (31), Ge (32), As (33), Se (34), Br (35), Kr (36), Sr (38), Zr (40), Mo (42), In (49), Sn (50), Sb (51), Te (52), I (53), Xe (54), and Ba (56). We can now add cesium ($Z = 55$) to this list as we have identified absorption lines of \ion{Cs}{iv}, \ion{Cs}{v}, and  \ion{Cs}{vi} in the {\it Far Ultraviolet Spectroscopic Explorer} (\textit{FUSE}) spectrum of the DO HD~149499B. However, determining the Cs abundance poses a challenge similar to that encountered by  T. Rauch, K. Werner, and collaborators in their analyses of trans-iron elements: the atomic data. Although \cite{sansonetti_2009} compiled identified energy level structures and transition wavelengths for the whole Cs isonuclear sequence, and \cite{hus20} reported a critical evaluation of energy levels, wavelengths, and $A$-values for \ion{Cs}{vii}, no transition probabilities for \ion{Cs}{iv}--\ion{Cs}{vi} have been measured or calculated to date. Consequently, our goal is to compute the respective oscillator strengths $f_{ik}$ and transition probabilities $A_{ki}$ for these species to determine the Cs abundance in the atmosphere of HD~149499B. Accurate radiative data are key to unravel the source of not only Cs but also of the other trans-iron elements observed in the atmospheres of hot white dwarfs. 


In this paper we recount the calculation of the transition probabilities for \ion{Cs}{iv}--\ion{Cs}{vi} and the ensuing determination of the Cs abundance in the atmosphere of HD~149499B. The \textit{FUSE} observations are described in Section~\ref{section:observations}. Section~\ref{section:atomic} summarises our approach to calculate the energy level structures and  transition probabilities for these ions. A description of the cesium abundance analysis by means of stellar atmosphere models is presented in Section~\ref{section:abundance_analysis}. We discuss our results and conclusions in Sections~\ref{section:discussion}--\ref{section:conclusion}.

\section{Far-Ultraviolet Spectroscopy}\label{section:observations}

\subsection{\textit{FUSE} Observations}\label{subsection:fuse}

\textit{FUSE} is a space-based telescope launched in 1999 as part of NASA's Origins program \citep{moos_etal2000apj} and deactivated in 2007. It is equipped with four independent spectrographs, two of which have LiF-coated optics sensitive to wavelengths between 990--1187~\AA, and two have SiC-coated optics sensitive to wavelengths between 905--1104~\AA.  The average resolution of the spectrographs is $R \simeq 20\,000$. Each spectrograph has a detector that records the signal on two independent microchannel-plate segments such that the spectral data are archived in eight files. The names of the segments and their wavelength coverage are: SiC1A (1004--1090 \AA), SiC1B (905--993 \AA), SiC2A (917--1005 \AA), SiC2B (1016--1104 \AA), LiF1A (988--1082 \AA), LiF1B (1094--1187 \AA), LiF2A (1087--1182 \AA), and LiF2B (979--1075 \AA).  

As HD~149499B was chosen as a reliable target to check the stability of the \textit{FUSE} photometric calibration, it was observed several times during the mission. Most of the observations were made under the photometric stability monitoring program M103 led by J.~W. Kruk. All the spectra observed under program M103 and through the $30\arcsec \times 30\arcsec$ aperture were retrieved from the Barbara A. Mikulski Archive for Space Telescope (MAST)\footnote{\url{https://mast.stsci.edu/portal/Mashup/Clients/Mast/Portal.html}}. In total, thirteen observations using this aperture were made between 2000 and 2006. Since HD149499B is a bright star in the far-ultraviolet wavelength range, all data were recorded in spectral-image mode (histogram: HIST).

The spectral data for each segment were first co-aligned and then co-added by taking the average of all exposures weighted by their exposure times. The SiC1B, LiF1A, SiC2B, LiF2A, and LiF1B segments are merged to form the final spectrum. Exposure times vary slightly from segment to segment depending on the quality of the individual exposures. The average total exposure time is approximately $66\,500$~s. Given the number of co-added spectra and the star's brightness, which is very close to the \textit{FUSE} bright limit, the signal-to-noise ratio of the resulting spectrum is of exceptional quality for a \textit{FUSE} spectrum. As the different segments have different sensitivities, the signal-to-noise ratio varies depending on the wavelength. It is equal to 65, 98, and 161 at wavelengths 950, 1050, and 1150~\AA. 

\subsection{Cesium detection}\label{subsection:cesium}

Considering the quality of the \textit{FUSE} spectrum, HD~149499B shows interstellar and stellar absorption lines with equivalent widths as narrow as 2 m\AA. \citet{wood_etal2002} measured the deuterium abundance and that of other species toward HD~149449B by analysing the interstellar lines detected in the \textit{FUSE} spectrum. They observed several absorption lines corresponding to ten species: \ion{H}{I}, \ion{D}{I}, \ion{C}{II}, \ion{C}{III}, \ion{N}{I}, \ion{N}{II}, \ion{O}{I}, \ion{P}{II}, \ion{Ar}{I}, and \ion{Fe}{II}. \citet{chayer_etal2005apjl} for their part analysed the stellar lines, reporting the detection of C, Si, P, S, and also of several trans-iron elements such as Ge, As, Se, Sn, Te, and I. They reported the detection of the \ion{Br}{VI} resonance line by using the wavelength published by \citet{morton_2000}, but it turned out, according to \citet{werner_etal2018_614}, that this wavelength was incorrect and was in fact longer by 6.7 \AA\ \citep{curtis_etal1984}. \citet{werner_etal2018_614} correctly identified the \ion{Br}{VI} line in question, additionally identifying two other \ion{Br}{VI} lines, and reported the first bromine detection in hot stars.

Although several interstellar and stellar species have been observed in the \textit{FUSE} spectrum, there are still several absorption lines that have not yet been identified. The identification of cesium lines from the \ion{Cs}{IV}, \ion{Cs}{V}, and \ion{Cs}{VI} species was possible by the publication of the observed spectral lines in the National Institute of Standards and Technology (NIST) Atomic Spectra Database\footnote{\url{https://www.nist.gov/pml/atomic-spectra-database}}  \citep{kra21}. Thirteen cesium lines have been identified in the \textit{FUSE} spectrum with equivalent widths in the range 2.3--26.9 m\AA. All these transitions arise from low-energy levels ranging from ten thousand to a few tens of thousands of cm$^{-1}$. These lines are listed in Table~\ref{tab:hd149499b_eqw}.

\begin{table*}
\centering
\begin{minipage}{160mm}
\caption{Atomic and observed properties and abundances of photospheric Cs lines identified in the \textit{FUSE} spectrum of HD~149499B. 
\label{tab:hd149499b_eqw}}
 \begin{tabular}{@{}lcccccccl@{}}
 \hline
Ion & Lab. ($\lambda$) & \multicolumn{2}{c}{$\log gf$} & $E_i$ & E.W. & \multicolumn{2}{c}{$\log N({\rm{Cs}})/N({\rm{He})}$} & Comments \\
\cline{3-4} \cline{7-8}
 & (\AA) & MCBP & MCDF & (cm$^{-1}$) & (m\AA) & MCBP & MCDF & \\
\hline
\ion{Cs}{IV} & 986.139  & $-1.022$ & $-1.557$ &  12902.0 & $2.3\pm0.7$ & $-6.09\pm0.41$ & $-5.54\pm0.41$ & very faint line \\
             & 1068.905 & $-1.355$ & $-2.084$ & 20754.0 & $2.3\pm0.3$ & $-5.90\pm0.50$ & $-5.15\pm0.50$ & very faint line\\
\ion{Cs}{V}  & 925.857  & $-1.840$ & $-1.770$ & 31951.1 & $8.7\pm0.8$ & $-5.10\pm0.37$ & $-5.23\pm0.37$ & next to ISM \ion{H}{i} $\lambda$926 line \\ 
             & 979.822  & $-0.942$ & $-1.022$ & 42273.7 & $15.1\pm0.5$ & $-5.99\pm0.33$ & $-5.92\pm0.33$ & \\
             & 1011.896 & $-1.253$ & $-2.059$ & 15077.4 & $23.9\pm0.5$ & $-5.92\pm0.34$ & $-5.10\pm0.34$ & \\
             &1064.881  & $-1.749$ & $-2.119$ & 31951.1 & $15.0\pm0.6$ & $-5.53\pm0.34$ & $-5.15\pm0.34$ & line appears broader than model \\
             &1069.196  & $-1.429$ & $-1.670$ & 20373.5 & $26.9\pm0.7$ & $\cdots$ & $\cdots$ & blend with \ion{Br}{v} $\lambda$1069 line\\
\ion{Cs}{VI} & 971.049  & $-0.871$ & $-1.039$ & 35061.4 & $21.4\pm0.9$ & $\cdots$ & $\cdots$ & blend with \ion{As}{iv} $\lambda$971 line\\
             & 982.436  & $-1.744$ & $-1.933$ & 52410.3 & $10.6\pm0.7$ & $-5.47\pm0.34$ & $-5.27\pm0.34$ & next to unidentified weak line\\
             & 1020.117 & $-2.348$ & $-2.592$ & 35061.4 & $4.8\pm0.3$ & $\cdots$ & $\cdots$ & uncertain identification\\
             & 1034.069 & $-1.619$ & $-1.712$ & 35061.4 & $9.5\pm0.6$ & $-5.47\pm0.34$ & $-5.38\pm0.34$ & \\
             & 1055.938 & $-1.621$ & $-1.887$ & 12176.0 & $23.8\pm0.5$ & $-5.31\pm0.31$ & $-5.04\pm0.31$ & \\
             & 1120.444 & $-1.621$ & $-1.830$ & 17628.2 & $19.9\pm0.4$ & $-5.33\pm0.32$ & $-5.11\pm0.32$ & next to \ion{I}{vi} $\lambda$1120 line and unidentified line\\
\hline
             &          &          &          &         & Average$^\dagger$: & $-5.61\pm0.35$ & $-5.29\pm0.35$ & \\
\hline
\end{tabular}
$^\dagger$ The average abundances can be expressed as mass fractions: $\log X_{\rm{Cs}} = -4.11$ for MCBP and $-3.79$ for MCDF. 
\end{minipage}
\end{table*}

\section{Cesium Atomic Data Calculations}\label{section:atomic}

The lack of oscillator strengths for  \ion{Cs}{IV}, \ion{Cs}{V}, and \ion{Cs}{VI} left us with no alternative but to calculate these data to estimate the cesium abundance in the atmosphere of HD149499B. The computation of the wave functions to determine $A$- and $gf$-values for atomic ions must take into account two important effects: electron correlation and the relativistic interactions. For multi-electron systems, the former is usually treated with configuration-interaction (CI) expansions \citep[see, for instance,][]{wei61}, which for heavy systems such as cesium with open $n\ell$ shells ($n\geq 5$ and $\ell\leq 3$) show very slow convergence. The relativistic interaction, on the other hand, can be approximated with a Pauli Hamiltonian (one-body operators), a Breit--Pauli Hamiltonian (one- and two-body operators), or a full Dirac Hamiltonian. For the heavy systems, attaining high overall accuracy while accounting simultaneously for these two effects is a daunting task; thus, the ionic representations we have adopted for the present work may only be regarded as a reconnaissance visit. We use the multiconfiguration Breit--Pauli (MCBP) method for the bulk of the calculations and the multiconfiguration Dirac--Fock (MCDF) method to get a measure of the accuracy of the radiative data, particularly for the observed lines.

\subsection{Multiconfiguration Breit--Pauli method}\label{section:mcbp}

Wave functions for the Cs ionic models were obtained with the multiconfiguration Breit--Pauli (MCBP) method through a CI expansion of the form
\begin{equation}\label{ci}
    \Psi(SLJ)=\sum_ic_i\psi(S_iL_iJ)
\end{equation}
using the \textsc{autostructure} atomic structure code \citep{eis74,bad11}. The basic configuration functions $\psi(S_iL_iJ)$ are built from $\phi(n\ell)$ one-electron orbitals obtained with the Thomas--Fermi--Dirac statistical model potential $V_\mathrm{mod}(\lambda_{n\ell})$ of \citet{eis69}, where the $\lambda_{n\ell}$ scaling parameters are variationally adjusted to minimize a weighted sum of the term energies of the ground and first-excited configurations. To obtain the radiative data, we performed modest calculations including only orbitals with principal quantum number $n\leq 7$ and orbital angular momentum $\ell\leq 2$. The configurations included in expansion~(\ref{ci}) for each Cs ion are tabulated in Table~\ref{tab:conf}.

\begin{table}
\centering
\caption{Electron configurations used in the present atomic models, where $n\ell=$ 5d, 6s, 6p, 6d, 7s. \label{tab:conf}}
 \begin{tabular}{@{}lll@{}}
 \hline
 \ion{Cs}{IV} & \ion{Cs}{V} & \ion{Cs}{VI}  \\
\hline
$\mathrm{5s^25p^4}$, $\mathrm{5s5p^5}$, $\mathrm{5p^6}$ & $\mathrm{5s^25p^3}$, $\mathrm{5s5p^4}$, $\mathrm{5p^5}$ &  $\mathrm{5s^25p^2}$, $\mathrm{5s5p^3}$, $\mathrm{5p^4}$ \\
$\mathrm{5s^25p^3}n\ell$ & $\mathrm{5s^25p^2}n\ell$ & $\mathrm{5s^25p}n\ell$ \\
$\mathrm{5s5p^4}n\ell$   & $\mathrm{5s5p^3}n\ell$   & $\mathrm{5s5p^2}n\ell$ \\
$\mathrm{5p^5}n\ell$     & $\mathrm{5p^4}n\ell$     & $\mathrm{5p^3}n\ell$   \\
$\mathrm{5s^25p^25d^2}$  & $\mathrm{5s^25p5d^2}$    & $\mathrm{5s^25d^2}$    \\
$\mathrm{5s^25p^25d6s}$  & $\mathrm{5s^25p5d6s}$    & $\mathrm{5s^25d6s}$    \\
$\mathrm{5s^25p^26s^2}$  & $\mathrm{5s^25p6s^2}$    & $\mathrm{5s^26s^2}$    \\
$\mathrm{5s5p^35d^2}$    & $\mathrm{5s5p^25d^2}$    & $\mathrm{5s5p5d^2}$    \\
$\mathrm{5s5p^35d6s}$    & $\mathrm{5s5p^25d6s}$    & $\mathrm{5s5p5d6s}$    \\
$\mathrm{5s5p^36s^2}$    & $\mathrm{5s5p^26s^2}$    & $\mathrm{5s5p6s^2}$    \\
$\mathrm{5p^45d^2}$      & $\mathrm{5p^35d^2}$      & $\mathrm{5p^25d^2}$    \\
$\mathrm{5p^45d6s}$      & $\mathrm{5p^35d6s}$      & $\mathrm{5p^25d6s}$    \\
$\mathrm{5p^46s^2}$      & $\mathrm{5p^36s^2}$      & $\mathrm{5p^26s^2}$    \\
\hline
\end{tabular}
\end{table}

The spectra of \ion{Cs}{vi}, \ion{Cs}{v}, and  \ion{Cs}{iv} listed in the NIST Atomic Spectra Database have been compiled by \citet{sansonetti_2009} from the measurements by \citet{tau91}, \citet{tau93}, and \citet{tau05} who used a normal incidence spectrograph with a triggered spark. Level identification was based on CI structure calculations with the \textsc{hfr} Pauli code \citep{cow81} and parametric least-squares fits. To reproduce these level structures and transition arrays, we have excluded configurations with 4f orbitals in the present atomic models.

The MCBP method relies on the relativistic Hamiltonian
\begin{equation}\label{ham}
    H(\mathrm{R})=H(\mathrm{NR})+H(\mathrm{BP})\ ,
\end{equation}
where $H(\mathrm{NR})$ is the usual non-relativistic Hamiltonian and $H(\mathrm{BP})$ contains the Breit—Pauli one- and two-body  relativistic corrections \citep{eis74}. Since Cs ions are large and complex, the CI expansions spanning the configurations listed in Table~\ref{tab:conf} are certainly non-convergent. To mitigate this shortcoming we introduce term energy corrections as described by \citet{men82}. The relativistic wave function $\psi_i(\mathrm{R})$ may be expanded to first order in terms of the non-relativistic counterparts $\psi_i(\mathrm{NR})$ as
\begin{equation}\label{tec}
   \psi_i(\mathrm{R}) = \psi_i(\mathrm{NR}) + \sum_{j\ne i}
   \psi_j(\mathrm{NR})\cdot\frac{\langle\psi_j(\mathrm{NR})|H(\mathrm{BP})|\psi_i(\mathrm{NR})\rangle}{\Delta E_{ij}(\mathrm{NR})}\ .
\end{equation}
The term energy corrections (TEC) 
\begin{equation}
\Delta E_\mathrm{nist}=\Delta E_{ij}(\mathrm{NR}) + \delta_i(\mathrm{NR})+\delta_j(\mathrm{NR}) 
\end{equation}
are introduced in the Hamiltonian (\ref{ham}) and adjusted empirically such that the denominator of the summation in Eq.~(\ref{tec}) is computed with the spectroscopic energy difference $\Delta E_\mathrm{nist}$ listed in the NIST database. 

\subsection{Multiconfiguration Dirac--Fock method}\label{section:mcdf}

To obtain a first estimate of $gf$ accuracy and a verification of level assignments, we performed a parallel calculation with the fully relativistic multi-configuration Dirac--Fock (MCDF) method \citep{gra80,mck80} using the same configuration expansion of Table~\ref{tab:conf}. The basic configuration functions  $\psi(S_iL_iJ)$ of Eq.~(\ref{ci}) are now expressed as antisymmetrized products of the orthonormal monoelectronic spin-orbitals 
\begin{equation}
\varphi_{n \kappa m} (r,\theta,\phi) = \frac{1}{r} \begin{pmatrix}
P_{n \kappa}(r)~\chi_{\kappa m}(\theta,\phi) \\
i~Q_{n \kappa} (r)~\chi_{-\kappa m}(\theta,\phi) \end{pmatrix}\ ,
\end{equation}
where $P_{n \kappa}(r)$ and $Q_{n \kappa}(r)$ are the large and small radial orbitals, respectively, and the angular functions $\chi_{\kappa m}(\theta,\phi)$ are spinor spherical harmonics. These spin-orbitals are optimized self-consistently based on the Dirac--Coulomb Hamiltonian
\begin{equation}
H_{DC}=\sum_i c \vec{\alpha_i} \cdot \vec{p_i}+ \beta_i c^2 - \frac{Z}{r_i}
+ \sum_{i>j} \frac{1}{r_{ij}}
\end{equation}
with the {\sc grasp2k} package \citep{par96}.

\subsection{Energy levels}\label{section:energies}

MCBP energy levels for \ion{Cs}{iv}, \ion{Cs}{v}, and  \ion{Cs}{vi} are respectively compared with the NIST spectroscopic measurements in Tables~\ref{tab:CsIV_energy_levels}, \ref{tab:CsV_energy_levels}, and \ref{tab:CsVI_energy_levels} of Appendix~\ref{appendix:A}. \ion{Cs}{vi} is the simpler spectrum with a $\mathrm{5s^25p^2}$ ground configuration (see Table~\ref{tab:CsVI_energy_levels}), whose energy differences with respect to NIST are within $\Delta E < 2000$~cm$^{-1}$; i.e., less than 1\%. The only inconsequential discrepancy is the configuration assignments of the strongly mixed $^1\mathrm{D}_2$ levels, for which we find the leading percentages $(\mathrm{5s^25p5d},\mathrm{5s5p^3})=(35,27)$ for the lowest level while NIST lists $(\mathrm{5s5p^3},\mathrm{5s^25p5d})=(32,28)$. As mentioned in Section~\ref{section:mcbp}, the NIST level assignments have been obtained with the \textsc{hfr} code, which in our experience are sometimes faulty.

The level comparison for \ion{Cs}{iv} ($\mathrm{5s^25p^4}$ ground configuration) in Table~\ref{tab:CsIV_energy_levels} is not as satisfactory: $\Delta E < 3000$~cm$^{-1}$ with respect to NIST exhibiting a few specific outliers:

\begin{itemize}
   \item $\mathrm{5s^25p^35d\ ^3D^o_2}$. We find $\Delta E > 14\,000$~cm$^{-1}$ for the two levels with this identification. These large discrepancies are difficult to explain and may be due to misquoted values by NIST.
   \item $\mathrm{5s^25p^35d\ ^3D^o_3}$. This level at  $E_\textsc{nist}=190\,309.9$~cm$^{-1}$ is very close ($\Delta E\approx 4600$~cm$^{-1}$) to $\mathrm{5s^25p^36s\ ^3D^o_3}$, and thus, the level assignments may be ambiguous: our level order is in fact the inverse of NIST.
   \item $\mathrm{5s^25p^35d\ ^3D^o_1}$. This level at  $E_\textsc{nist}=189\,391.7$~cm$^{-1}$ is very close ($\Delta E\approx 3500$~cm$^{-1}$) to $\mathrm{5s^25p^35d\ ^3P^o_1}$ whereby both levels showing strong admixture.
\end{itemize}

Regarding the energy level structure of \ion{Cs}{v} ($\mathrm{5s^25p^3}$ ground configuration), the situation is more complicated as the relativistic level admixture is stronger than in the two previously discussed ions, and as shown in the additional column of Table~\ref{tab:CsV_energy_levels}, our CI expansions suggest reassignments for the following NIST levels:

\begin{itemize}
 \item $\mathrm{5s5p^4\ ^2S_{1/2}}$, $\mathrm{5s5p^4\ ^2P_{1/2}}$, and $\mathrm{5s^25p^25d\ ^2P_{1/2}}$
 \item $\mathrm{5s^25p^25d\ ^4D_{5/2}}$ and $\mathrm{5s^25p^25d\ ^2F_{5/2}}$
 \item $\mathrm{5s^25p^25d\ ^4D_{7/2}}$ and $\mathrm{5s^25p^25d\ ^2F_{7/2}}$
 \item $\mathrm{5s^25p^26s\ ^2P_{3/2}}$ and $\mathrm{5s^25p^25d\ ^2P_{3/2}}$ 
 \item $\mathrm{5s^25p^26s\ ^2P_{1/2}}$ and $\mathrm{5s^25p^25d\ ^2S_{1/2}}$.  
\end{itemize}
Once these changes in level identifications are made, the energy differences between our computed energies and those of NIST are $\Delta E \lesssim 3000$~cm$^{-1}$. Furthermore, \citet{huo17} have used the MCDF {\sc grasp2k} package to compute level energies for the ground configuration of \ion{Cs}{v}. With the exception of the $\mathrm{5s^25p^3\ ^2P^o_{3/2}}$ level, our MCBP values are marginally closer to NIST due to the fine-tuning procedure based on TEC (see Section~\ref{section:mcbp}).

\subsection{Wavelengths}\label{section:wl}

Radiative data for the allowed transition arrays of the three ions computed with the MCBP method are listed in Tables~\ref{tab:CsIV_list_lines}--\ref{tab:CsVI_allowed_transitions}. Regarding computed transition wavelengths, the best we can do is exemplified with \ion{Cs}{vi} (see Table~\ref{tab:CsVI_allowed_transitions}), where the differences with the NIST wavelengths are on average $\overline{\Delta\lambda} = {-0.1}\pm 2.4$~\AA. The larger differences, $4.0\leq \Delta\lambda\leq 8.5$~\AA, occur in transitions with wavelengths $\lambda > 1000$~\AA\  involving the  $\mathrm{5s^25p^2\ ^1D_2}$ lower level. 

For \ion{Cs}{iv}, on the other hand, the wavelength comparison with NIST is poorer: $\overline{\Delta\lambda} = {+0.1}\pm 15.1$~\AA. However, if we exclude the transitions involving the two $\mathrm{5s^25p^35d\ ^3D^o_2}$ levels discussed in Section~\ref{section:energies}, then the average difference is reduced to a more acceptable $\overline{\Delta\lambda} = {+0.5}\pm 7.0$~\AA. Thus, in Table~\ref{tab:CsIV_allowed_transitions} we question the following NIST wavelengths:  558.161~\AA\ and 601.476~\AA\ of transition array 11 (TA-11) ; 586.848~\AA\ and 634.921~\AA\ of TA-17; 631.290~\AA\ of TA-35 and 589.776 \AA\ of TA-41. 

The \ion{Cs}{v} average wavelength difference with NIST is  remarkable: $\overline{\Delta\lambda} = {-2.3}\pm 7.1$~\AA. Although the standard deviation is not larger than in \ion{Cs}{iv}, the average difference indicates that the theoretical wavelength as a whole are shorter by ${\sim}2$~\AA. This is a symptom of the difficulties encountered when trying to fit a spectroscopic spectrum with TEC under pervasive level coupling.

\subsection{Radiative rates and $f$-values}\label{section:fval}

Tables~\ref{tab:CsIV_allowed_transitions}, \ref{tab:CsV_allowed_transitions}, and \ref{tab:CsVI_allowed_transitions} tabulate MCBP $A$- and $f$-values for the allowed (E1) lines, and to the best of our knowledge, it is the first time that radiative rates are reported for these ionic species apart from the work by \cite{bie95} on the forbidden (E2 and M1) lines within the ground configuration. Nonetheless, semi-empirical branching fractions have been previously estimated for the $\mathrm{5s^25p^2}-\mathrm{5s^25p6s}$ transition array by \citet{cur01}. The $F_2$ and $G_1$ Slater parameters and $\zeta_\mathrm{p}$ and $\zeta_\mathrm{pp}$ spin--orbit energies where obtained by fitting the spectroscopic level energies to then derive singlet--triplet mixing angles and, hence, relative transition rates. In Table~\ref{bf} we show a comparison of these branching fractions with those obtained from the present MCBP $A$-values. Regarding allowed lines with $\Delta S=0$, the agreement for the $\mathrm{^3P^o-{^3P}}$ transitions is ${\sim}20\%$ while significant discrepancies appear for $\mathrm{^1P^o_1-{^1D}_{2}}$ and $\mathrm{^1P^o_1-{^1S_{0}}}$, particularly a factor of 3 for the latter. For the intercombination transitions  ($\Delta S\ne 0$), factor differences are found for $\mathrm{^3P^o_1-{^1S}_{0}}$ and $\mathrm{^1P^o_1-{^3P}_{1}}$.

Moreover, we compare the MCBP and MCDF $gf$-values for the observed Cs lines in Table~\ref{tab:hd149499b_eqw}. The agreement is, as expected, relatively poor: within 0.3~dex apart from the three lines $\lambda\lambda$986.139, 1068.905, 1011.896 that display even larger discrepancies. The latter may be due to divergent CI expansions, strong admixture, and cancellation effects.

\begin{table}
\centering
\caption{Comparison of branching fractions for the $\mathrm{5s^25p6s-5s^25p^2}$ transition array of \ion{Cs}{vi}. BFP: present. BFC: \citet{cur01}. \label{bf}}
 \begin{tabular}{@{}lrcrr@{}}
 \hline
 Upper level & Lower level & $\lambda$ & BFP & BFC \\
             &             & (\AA)    & (\%) & (\%) \\
\hline
$\mathrm{5p6s\ ^3P^o_1}$ & $\mathrm{5p^2\ ^3P_0}$ & 410.31 & 26.4 & 31.0 \\ 
                         & $\mathrm{^3P_1}$       & 431.89 & 15.1 & 15.6 \\
                         & $\mathrm{^3P_2}$       & 442.30 & 41.4 & 51.2 \\
                         & $\mathrm{^1D_2}$       & 479.25 &  2.0 &  1.8 \\
                         & $\mathrm{^1S_0}$       & 522.72 & 16.9 &  0.3 \\
$\mathrm{5p6s\ ^3P^o_2}$ & $\mathrm{5p^2\ ^3P_1}$ & 401.97 & 24.5 & 27.7 \\ 
                         & $\mathrm{^3P_2}$       & 410.98 & 46.2 & 51.1 \\
                         & $\mathrm{^1D_2}$       & 442.69 & 29.3 & 21.2 \\
$\mathrm{5p6s\ ^1P^o_1}$ & $\mathrm{5p^2\ ^3P_0}$ & 378.46 &  0.1 &  0.6 \\ 
                         & $\mathrm{^3P_1}$       & 396.75 &  1.4 &  4.4 \\
                         & $\mathrm{^3P_2}$       & 405.52 &  9.0 & 11.9 \\
                         & $\mathrm{^1D_2}$       & 436.37 & 49.4 & 68.6 \\
                         & $\mathrm{^1S_0}$       & 472.11 & 40.1 & 14.6 \\                         
\hline
\end{tabular}
\end{table}

\section{Cesium Abundance Analysis}\label{section:abundance_analysis}

We determine the cesium content of the star's atmosphere by comparing the Cs lines observed in the HD~149499B \textit{FUSE} spectrum with a grid of synthetic spectra.  The grid is calculated for different Cs abundances and for a given effective temperature, gravity, and $N({\rm{H}})/N({\rm{He}})$ ratio. The synthetic spectra are themselves calculated from a grid of stellar atmosphere models that specify the physical properties of the upper layer of the star.  We use the stellar atmosphere and spectrum synthesis codes {\tt Tlusty}\footnote{\url{http://tlusty.oca.eu}}  and {\tt Synspec}\footnote{\url{http://tlusty.oca.eu/Synspec49/synspec.html}} to compute non-local thermodynamic equilibrium (non-LTE) stellar atmosphere models and synthetic spectra of HD~149499B.  A description of the two codes and their use is given in a series of three papers by \citet{hubeny_lanz_2017arxiv_a, hubeny_lanz_2017arxiv_b, hubeny_lanz_2017arxiv_c}.

\subsection{Model atmospheres}\label{subsection:stellarmodels}

As HD~149499B has a bright companion separated by only 2 arcsecs, no reliable optical spectroscopic study has thus far been performed  \citep{dreizler_werner96}. We adopt the atmospheric parameters determined by \citet{napiwotzki_etal1995} in the analysis of the far-ultraviolet spectrum observed with the Berkeley spectrograph aboard the ORFEUS\footnote{Orbiting Retrievable Far and Extreme Ultraviolet Spectrometers} telescope. This spectrograph covers a wavelength range between 390 and 1170~\AA\ with a resolution $R = \lambda/3000$. They fitted the stellar \ion{H}{I} and \ion{He}{II} lines observed beyond the Lyman edge at 912~\AA\ with a grid of local thermodynamic equilibrium (LTE) stellar atmosphere models to determine the following stellar parameters: $T_{\rm{eff}} = 49\ 500\pm500$~K, $\log g = 7.97\pm0.08$, and $\log N({\rm{H}})/N({\rm{He}}) = -0.65\pm0.12$. \citet{jordan_etal97} confirmed this temperature by analysing spectra collected by the {\it Extreme Ultraviolet Explorer}. By fitting the medium-wavelength (140--380 \AA) and long-wavelength (280--760 \AA) spectra, they estimated a temperature identical to that by \citet{napiwotzki_etal1995} and an interstellar hydrogen column dentity $N_{\rm{H}} = 7\times 10^{18}$cm$^{-2}$ along the sight line of HD~149499B. 

The model H and He atoms considered in our model atmosphere calculations are the same as those described by \citet{lanz_hubeny_2003, lanz_hubeny_2007apjs}, who calculated grids of non-LTE line-blanketed model atmospheres of B- and O-type stars. These model atoms are available from the {\tt Tlusty} website\footnote{\url{http://tlusty.oca.eu/Tlusty2002/tlusty-frames-data.html}}. Based on the atomic data calculated with the MCBP method described in Section~\ref{section:mcbp} and those compiled by \citet{sansonetti_2009}, we built the cesium model atom following the description of \citet{hubeny_lanz_2017arxiv_b, hubeny_lanz_2017arxiv_c}. There is a model atom per cesium ion. We consider the \ion{Cs}{iv}, \ion{Cs}{v}, \ion{Cs}{vi}, and \ion{Cs}{vii} species. Each model atom consists of a set of energy levels and bound--free and bound--bound transitions. In order to keep a relatively small number of energy levels in the calculations, only the terms given in Tables \ref{tab:CsIV_energy_levels},  \ref{tab:CsV_energy_levels}, and \ref{tab:CsVI_energy_levels} are considered. This brings the number of energy levels to respectively 29, 22, and 17 for the \ion{Cs}{iv}, \ion{Cs}{v}, and \ion{Cs}{vi} ions. \ion{Cs}{vii} is considered as a single-energy level ion. Radiative bound--free transitions are described by hydrogenic cross sections, while collisional bound--free transitions are expressed in terms of photoionisation cross sections by the so-called \citet{seaton62} formula. Allowed bound--bound transitions are described by depth-independent Doppler line profiles with a temperature given by $T = 0.75T_{\rm{eff}}$. Total multiplet oscillator strengths $f_{ik}$ are given in Tables \ref{tab:CsIV_allowed_transitions}, \ref{tab:CsV_allowed_transitions}, and \ref{tab:CsVI_allowed_transitions}. For allowed line transitions, collisional excitation cross sections are computed using the \citet{vanregemorter1962} formula, and for the forbidden line transitions, the cross sections are obtained by the \citet{eissner_seaton1972} formula with $\gamma(T) = 0.5$.

\begin{figure}
\includegraphics[width=\columnwidth]{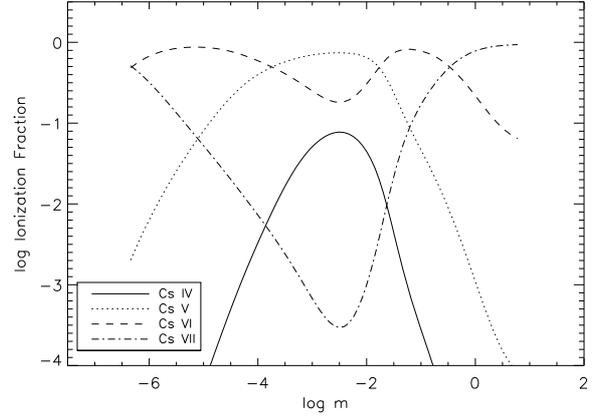}
\caption{Ionization fractions of \ion{Cs}{iv}--\ion{Cs}{vii} as a function of the column mass $m$ (g~cm$^{-2}$) in a NLTE model of HD~149499B calculated with the atmospheric parameters provided by \citet{napiwotzki_etal1995}: $T_{\rm{eff}} = 49$,500~K, $\log g = 7.97$, and $\log N({\rm{H}})/N({\rm{He}}) = -0.65$. }
\label{fig:cs_ionization_frac}
\end{figure}

We computed non-LTE stellar atmosphere models with a H, He, and Cs chemical composition. We considered the range of Cs abundances $\log N({\rm{Cs}})/N({\rm{He}}) = -9.6$ to $-4.4$ in steps of 0.4. We also computed models that take into account the uncertainties of the atmospheric parameters ($\Delta T_{\rm{eff}}$, $\Delta \log g$, $\Delta \log N({\rm{H}})/N({\rm{He}})$). The uncertainty on the effective temperature was increased to 2000~K to improve the assessment of the Cs abundance change with the effective temperature. Since the atmosphere of HD~149499B is rich in helium, the models were calculated by selecting the parameter ${\rm{IATREF}} = 2$, which implies that helium was chosen as the reference atom. Fig.\ref{fig:cs_ionization_frac} compares the ionisation fractions of Cs in a model of HD~149499B showing that \ion{Cs}{v} and \ion{Cs}{vi} are the dominant ions in the line-forming region of the atmosphere, which is located between $\log m \approx -1.5$ and $-4.0$. These ionisation fractions agree well with the detection of the \ion{Cs}{V} and \ion{Cs}{VI} lines in the \textit{FUSE} spectrum of HD~149499B. The detection of weak \ion{Cs}{IV} lines, which have $\log gf$ comparable to those of the \ion{Cs}{V} and \ion{Cs}{VI}, also illustrates that the ionisation fraction of \ion{Cs}{IV} is about an order of magnitude lower than that of \ion{Cs}{V}.

\begin{figure*}
\includegraphics[width=160mm]{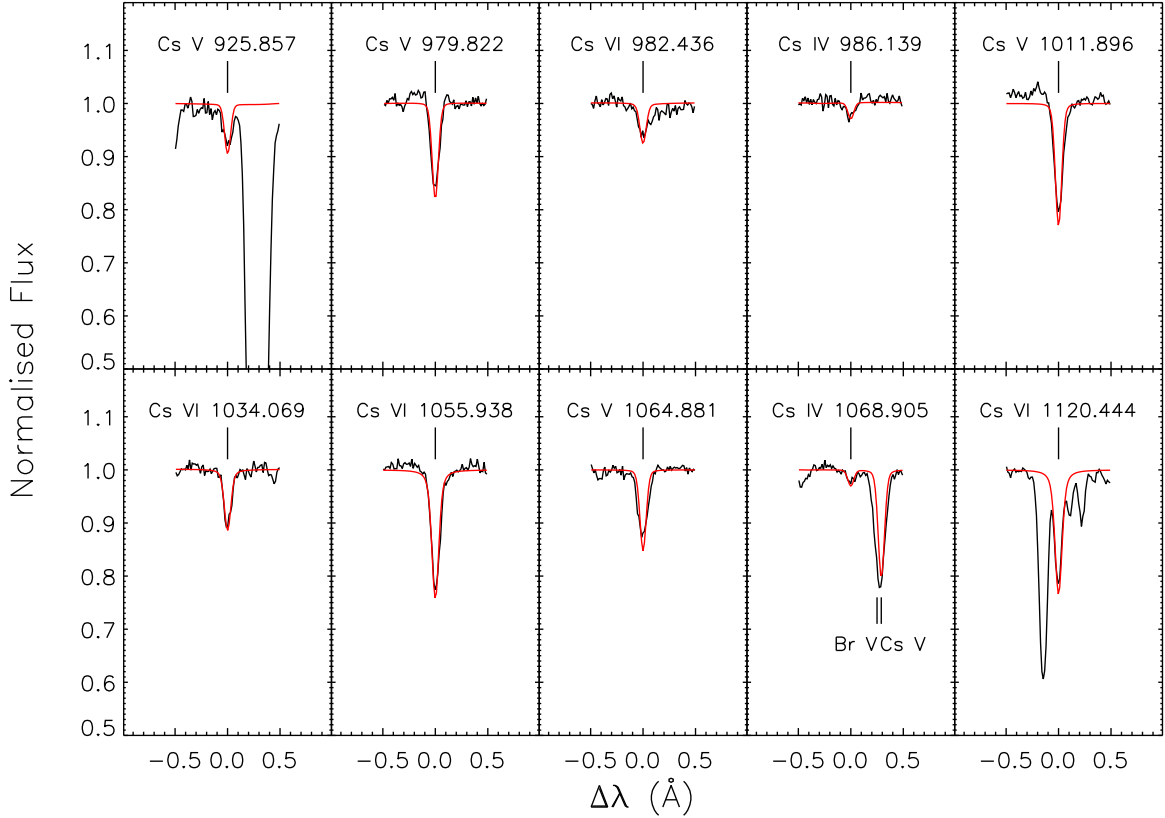}
\caption{Best-fitting (red curves) to the \ion{Cs}{iv}--\ion{Cs}{vi} lines observed in the \textit{FUSE} spectrum of HD~149499B (black curves). The fits are computed by considering the $\log gf$ obtained by the MCBP method. Abundances are given in Table~\ref{tab:hd149499b_eqw}. }\label{fig:cs_line_fits}
\end{figure*}

\subsection{Spectral fitting}\label{subsection:fitting}

The Cs abundance is determined by comparing the cesium lines in the synthetic spectra to those observed in the \textit{FUSE} spectrum. Synthetic spectra covering the \textit{FUSE} wavelength range 905--1187~\AA\ are calculated by using the spectral synthesis code {\tt Synspec}. This code uses the models of stellar atmospheres calculated by {\tt Tlusty} as input. The line list was retrieved from the {\tt Synspe}c website\footnote{\url{http://tlusty.oca.eu/Synspec49/synspec-frames-data.html}} and supplemented with the Cs lines given in Tables~\ref{tab:CsIV_allowed_transitions}, \ref{tab:CsV_allowed_transitions}, and \ref{tab:CsVI_allowed_transitions}. The main {\tt Synspec} output is the emergent flux listed in a table with the wavelength (\AA) and the flux expressed in terms of the flux moment, $H_\lambda$, in units of erg~cm$^{-2}$~s$^{-1}$~\AA$^{-1}$. The synthetic spectra are convolved with a Gaussian of {\sc FWHM} = 0.06 \AA\ such that their resolution matches that of \textit{FUSE}; they are then normalised to the theoretical continuum provided by {\tt Synspec}. Before making a comparison with the synthetic spectra, the observed lines are normalised and shifted to the laboratory rest frame.  Using a fitting routine, the Cs abundance and its uncertainty are determined when the theoretical and observed lines match. The fitting routine calls a function that linearly interpolates among the synthetic spectra and fits them to each line individually by taking the abundance as a free parameter. Fig.~\ref{fig:cs_line_fits} shows the best fits to the \ion{Cs}{iv}--\ion{Cs}{vi} lines. 

\subsection{Results: Cesium abundance}\label{subsection:results}

Table~\ref{tab:hd149499b_eqw} lists the atomic and observed properties and abundances of the \ion{Cs}{iv}--\ion{Cs}{vi} lines detected in the \textit{FUSE} spectrum. As described in Section~\ref{section:observations} and illustrated in Fig.~\ref{fig:cs_line_fits}, the quality of the co-added \textit{FUSE} spectrum allows the detection of very weak lines. For instance, the two detected \ion{Cs}{iv} lines have equivalent widths of only 2.3 m\AA, and although these lines are very weak, they are clearly visible as shown in Fig.~\ref{fig:cs_line_fits}. A pair of lines are blended with known lines: \ion{Cs}{v} $\lambda$1069 and \ion{Cs}{vi} $\lambda$971 with the \ion{Br}{v} $\lambda$1069 and \ion{As}{iv} $\lambda$971, respectively. The \ion{Cs}{vi} $\lambda$1020 line, on the other hand, appears to be blended with unidentified faint lines that slightly depress the continuum in the blue wing of the Lyman~$\beta$ and \ion{He}{ii} $\lambda$1024 lines. No Cs abundance was derived from these three lines. 

Table~\ref{tab:hd149499b_eqw} compares the oscillator strengths calculated with the {\sc MCBP} and {\sc MCDF} methods; it also gives the Cs abundances determined with these two datasets.  The MCBP oscillator strengths are systematically larger than those calculated with MCDF. This results in abundances that are lower by a factor of two on average for the MCBP method. In both cases, the standard deviation of the abundance measurements for the ten lines is about 0.3 dex. The abundance uncertainties take into account those on the atmospheric parameters and the oscillator strengths and the quality of the fit. The uncertainties on the atmospheric parameters we have considered are $\Delta T_{\rm{eff}} = 2\ 000$~K, $\Delta \log g = 0.08$, and $\Delta \log N({\rm{H}})/N({\rm{He}}) = 0.12$. The uncertainties on the oscillator strengths are estimated at 50\%. On average, the contributions of these uncertainties to that on the abundance are: 0.16 dex for the effective temperature; 0.04 dex for the gravitational acceleration; 0.01 dex for the H abundance; 0.31 dex for the oscillator strength; and 0.04 dex for the statistical errors. The contributions of these uncertainties to the abundance uncertainty are combined in quadrature. The average abundances are $\log N({\rm{Cs}})/N({\rm{He}}) = -5.61\pm0.35$ and $-5.29\pm0.35$ for the MCBP and MCDF methods, respectively. The corresponding mass fractions are $\log X_{\rm{Cs}} = -4.11\pm0.35$ and $-3.79\pm0.35$. Taking the average of both methods, the Cs abundance and its mass fraction then become $-5.45\pm0.35$ and $-3.95\pm0.35$.

\section{Discussion}\label{section:discussion}

\subsection{Presence of cesium}\label{subsection:abundance}

Cesium detection in the atmosphere of HD~149449B is not surprising, because the list of trans-iron elements detected in the atmospheres of white dwarfs has grown considerably since that of germanium by \cite{vennes_etal2005apj}. Cesium is now the nineteenth trans-iron element detected in the atmosphere of a white dwarf. Like other trans-iron elements observed in hot white dwarfs, cesium is overabundant by approximately 125,000 times solar. It is the most abundant trans-iron element in HD~149499B followed by bromine, iodine, selenium, arsenic, tin, tellurium, and germanium \citep{chayer_etal2005apjl, werner_etal2018_614}. This cesium abundance is comparable, for instance, to other trans-iron elements observed in the hotter DO RE~0503$-$289 \citep{hoyer_etal2018}.  


The presence of cesium and other trans-iron elements in the atmospheres of hot white dwarfs is remarkable. As white dwarfs are well known for their intense gravitational fields, a physical mechanism must counteract the effects of downward element diffusion to maintain these trans-iron elements in the atmosphere. In hot white dwarfs, radiative levitation is the natural mechanism to explain their presence \citep{vauclair_etal1979aa, chayer_etal1995a, dreizler_wolff1999aa, rauch_etal2016_587}. The radiation field in these hot stars provides enough momentum through bound--bound absorption to allow the heavy elements to levitate in the atmosphere. As the momentum transferred to an element depends on its abundance,  determination of the latter is possible when the radiative acceleration equals the gravitational acceleration. Based on the equilibrium radiative levitation theory, \citet{rauch_etal2016_587} calculated equilibrium distributions of nine trans-iron elements at the surfaces of the DA G191$-$B2B and DO RE~0503$-$289. They demonstrated that radiative levitation can support trans-iron elements with equilibrium abundances showing overabundances in the line-forming regions. In future work, we intend to show that radiative levitation can maintain cesium in the atmosphere of HD~149499B.


\subsection{Source of cesium}\label{subsection:source}


Could the presence of cesium in the atmosphere of HD~149499B indicate that lighter elements have undergone exposure to slow neutron capture ($s$-process) during previous phases of the star's evolution? Models describing the evolution of asymptotic-giant-branch (AGB) stars show that $s$-process nucleosynthesis of trans-iron elements is possible for low- and intermediate-mass stars \citep{iben_renzini_1983, busso_etal1999, herwig_2005}. The detection of these elements in the atmospheres of white dwarfs could then provide clues to understand better the role of $s$-process nucleosynthesis during the AGB phase. However, because diffusion processes such as gravitational settling and radiative levitation erase all traces of the abundance history, it is not possible to disentangle the trans-iron element abundances produced by $s$-process nucleosynthesis during the AGB phase from those present during star formation. Although radiative levitation complicates the interpretation of the source of trans-iron elements, it enables the buildup of large abundances, and therefore, the detection of those elements that otherwise would not be detected. As suggested by \cite{werner_etal2015_574}, confirmation of $s$-process nucleosynthesis would entail the detection of a radioactive element such as technetium (Tc). The detection of this element in the atmosphere of red-giant stars by \cite{merrill_1952} confirmed that $s$-process nucleosynthesis had taken place in evolved stars. In the case of hot white dwarfs, radiative levitation would favor the detection of technetium even if it disintegrates over time.  Technetium would diffuse from deeper regions in the atmosphere to the surface by radiative levitation, and would show an abundance when the radiative acceleration equals the gravitational acceleration. Ultimately, technetium would disappear from the atmosphere, but its disappearance would be delayed by radiative levitation. 

\subsection{Cesium atomic data}\label{subsection:atomic_data}

Computing radiative data for heavy ions such as \ion{Cs}{iv}--\ion{Cs}{vi} has been an interesting challenge. Ionic species with ground configurations $n\mathrm{s}^2n\mathrm{p}^m$ ($2\leq n\leq 3$ and $2\leq m\leq 4$) are familiar in nebular astrophysics, but the required $A$-values are mostly for line-ratio diagnostics involving forbidden lines. In the present study, $n=5$, but conveniently, the level structures display similar symmetries simplifying the atomic representation strategies. However, for hot white-dwarf atmospheric models, the required radiative rates are for allowed transition arrays involving levels from the lowly excited configurations displaying 5d and 6s orbitals. Slow CI expansion convergence and strong relativistic couplings are then more difficult to account for to ensure acceptable accuracy.  

We were surprised by the lack of previous $A$-value estimates despite the comprehensive Cs spectrum compilation by \citet{sansonetti_2009}, which includes well-identified spectroscopic level structures for \ion{Cs}{iv}--\ion{Cs}{vi} by \citet{tau91,tau05} and \citet{tau93}. Such level identifications were based on \textsc{hfr} CI calculations, which for configurations of the type $n\mathrm{s}n\mathrm{p}^m$ ($3\leq m\leq 5$) can be faulty as reported by \citet{pal12}. The present level assignments resulting from adjusting the theoretical term energies to the spectroscopic values with TEC are expected to be more reliable. A problem that emerges in this procedure is that, due to strong relativistic couplings and level admixture, term splittings can be large; thus, $LS$-averaged wavelengths will not necessarily lead to adequate representations of the observed transition arrays.

Without comparisons with independent datasets, it is difficult to give an estimate of the accuracy level of the present radiative rates. In our attempt to reproduce the observed spectra, we kept our CI expansions as concise as possible, excluding for instance configurations with 4f orbitals. It is difficult to predict a priori whether this approximation leads to significant limitations. The comparison with the semi-empirical branching ratios for the $\mathrm{5s^25p^2}-\mathrm{5s^25p6s}$ transition array by \citet{cur01} showed nonetheless significant discrepancies that can be attributed to relativistic effects. We intend as a followup to the present report to recompute the present radiative datasets with a multiconfiguration Dirac--Fock method in an attempt to constrain accuracy ratings and to determine collision strengths and photoinisation cross sections with \textsc{autostructure} to improve the non-LTE atmospheric models. Therefore, from the atomic physics point of view, the present Cs abundance determinations is open to improvement. 

\section{Conclusion}\label{section:conclusion}

The detection of cesium in the atmosphere of HD~149499B is the latest addition to a list of trans-iron elements observed in the atmospheres of hot white dwarfs. Absorption lines from \ion{Cs}{iv}, \ion{Cs}{v}, and \ion{Cs}{vi} are present in the \textit{FUSE} spectrum. First, we determined the oscillator strengths of the bound--bound transitions of these three ions using state-of-the-art atomic structure codes. We then determined the cesium abundance using stellar atmosphere models. With a mass fraction of $\log X_\text{Cs} = -3.95\pm0.35$, cesium is the most abundant trans-iron element observed in HD~149499B. Radiative levitation is the most plausible natural phenomenon to explain its presence. Although radiative levitation favors its observation, it complicates the search of its origin. 


\section*{Acknowledgements}
This research has made use of the SIMBAD database, operated at CDS, Strasbourg, France. All of the astronomical observation data presented in this paper were obtained from the Mikulski Archive for Space Telescope (MAST). STScI is operated by the Association of Universities for Research in Astronomy, Inc., under NASA contract NAS5-26555. Support for MAST for non-HST data is provided by the NASA Office of Space Science via grant NNX13AC07G and by other grants and contracts. This research has made use of NASA's Astrophysics Data System. 

\section*{DATA AVAILABILITY}
The data underlying this article will be shared on reasonable request to the corresponding author.




\bibliographystyle{mnras}
\bibliography{wd_bib}



\appendix

\section{Cesium Atomic Data}\label{appendix:A}

The present determination of the Cs abundance in the hot white dwarf HD~149499B required fairly extensive calculations of the level structures and radiative rates of \ion{Cs}{IV}, \ion{Cs}{V}, and \ion{Cs}{VI}. In the following tables, we tabulate MCBP energy levels and $A$-, $f$-, and $\log(gf)$-values for allowed transitions in these ionic species. MCBP and NIST energy levels for \ion{Cs}{IV}, \ion{Cs}{V}, and \ion{Cs}{VI} are compared in Tables~\ref{tab:CsIV_energy_levels}, \ref{tab:CsV_energy_levels}, and \ref{tab:CsVI_energy_levels}. Atomic transition probabilities are given in Tables~\ref{tab:CsIV_allowed_transitions}, \ref{tab:CsV_allowed_transitions}, and \ref{tab:CsVI_allowed_transitions}. Finding lists ordered in increasing wavelengths are given in Tables \ref{tab:CsIV_list_lines},  \ref{tab:CsV_list_lines}, and \ref{tab:CsVI_list_lines}. These finding lists correspond to the atomic transitions that are reported in Tables~\ref{tab:CsIV_allowed_transitions}, \ref{tab:CsV_allowed_transitions}, and \ref{tab:CsVI_allowed_transitions}, respectively.



\begin{table*}
\centering
\begin{minipage}{140mm}
\caption{\ion{Cs}{IV}: Comparison of MCBP and NIST energy levels. \label{tab:CsIV_energy_levels}}
\begin{tabular}{lccrrlccrr}
\hline
Configuration & Term & $J$ & $E$(Present) & $E$(NIST) &  Configuration & Term & $J$ & $E$(Present) & $E$(NIST) \\     
              &      &     & (cm$^{-1}$) & (cm$^{-1}$) & &     &     & (cm$^{-1}$) & (cm$^{-1}$) \\
\hline
$\mathrm{5s^25p^4}$  &  $^3\mathrm{P}$	      & 2 & 0.0	      & 0.0	         & $\mathrm{5s^25p^35d}$ & $^3\mathrm{D}^\mathrm{o}$ & 1 & 159501.2 & 162425.9 \\       
                       & $^3\mathrm{P}$	              & 0 & 10001.8   & 9749.4	&  	               & $^3\mathrm{D}^\mathrm{o}$                     & 2 & 165031.1 & 179159.8 \\  
                       & $^3\mathrm{P}$	              & 1 & 13353.5   & 12902.0   & 	               & $^3\mathrm{D}^\mathrm{o}$                     & 3 & 170539.7 & 173942.4 \\
                       &  term                                      &    & 5562.5     & 5383.9       &                     & term                                                           &    & 166495.8 & 173378.2 \\
$\mathrm{5s^25p^4}$  & 	$^1\mathrm{D}$     & 2 & 21069.1   & 20754.0	& $\mathrm{5s^25p^35d}$  & $^3\mathrm{F}^\mathrm{o}$ & 3 & 175067.3 & 174789.4 \\ 
                       &  term                                      &    & 21069.1   & 20754.0   &                      & $^3\mathrm{F}^\mathrm{o}$                     & 2 & 175677.9 & 174718.7 \\
\hline
\end{tabular}
\end{minipage}
\end{table*}



\begin{table*}
\centering
\begin{minipage}{140mm}
\caption{\ion{Cs}{V}: Comparison of MCBP and NIST energy levels. In column $E_\mathrm{ra}$(NIST) several NIST levels have been reassigned. \label{tab:CsV_energy_levels}}
\begin{tabular}{lccrrrlccrrr}
\hline
Conf & Term & $J$ & $E$(Pres) & $E_{\rm ra}$(NIST) & $E$(NIST) & Conf & Term & $J$ & $E$(Pres) & $E_{\rm ra}$(NIST) & $E$(NIST) \\     
& & & (cm$^{-1}$) & (cm$^{-1}$) & (cm$^{-1}$) & & & & (cm$^{-1}$) & (cm$^{-1}$) & (cm$^{-1}$) \\
\hline
$\mathrm{5s^25p^3}$	    &	$^4\mathrm{S^o}$	& 3/2 &	0.0	        &	0.0	        &	0.0	       & $\mathrm{5s^25p^25d}$	&	$^2\mathrm{G}$	& 7/2 &	190802.0	&	189415.9   &	189415.9	\\                          
$\mathrm{5s^25p^3}$	    &	$^2\mathrm{D^o}$	& 3/2 &	15025.3	    &	15077.4	    &	15077.4    &     &	$^2\mathrm{G}$	& 9/2 &	193152.4	&	192876.3   &	192876.3	\\                           
	       &	$^2\mathrm{D^o}$	& 5/2 &	20408.2	    &	20373.5	    &	20373.5  &    	                    &	term	        &     &	192107.8	&	191338.3	&	191338.3	\\                             
	       &	term	        &	  & 18255.0	    &	18255.1	    &	18255.1  &        $\mathrm{5s5p^4}$	    &	$^2\mathrm{P}$	& 3/2 &	191078.9	&	188245.8	&	188245.8	\\                            
$\mathrm{5s^25p^3}$	    &	$^2\mathrm{P^o}$	& 1/2 &	31717.5	    &	31951.1	    &	31951.1	 & 	                    &	$^2\mathrm{P}$	& 1/2 &	202678.7	&	201141.6	&	158248.8	\\                           
	                    &	$^2\mathrm{P^o}$	& 3/2 &	42390.5	    &	42273.7	    &	42273.7  &    	                    &	term	        &     &	194945.5	&	192544.4	&	178246.8	\\
\hline
\end{tabular}
\end{minipage}
\end{table*}



\begin{table*}
\centering
\begin{minipage}{140mm}
\caption{\ion{Cs}{VI}: Comparison of MCBP and NIST energy levels. \label{tab:CsVI_energy_levels}}
\begin{tabular}{lccrrlccrr}
\hline
Configuration & Term & $J$ & $E$(Present) & $E$(NIST) &  Configuration & Term & $J$ & $E$(Present) & $E$(NIST) \\     
              &      &     & (cm$^{-1}$) & (cm$^{-1}$) & &     &     & (cm$^{-1}$) & (cm$^{-1}$) \\
\hline
$\mathrm{5s^25p^2}$  & 	$^3\mathrm{P}$	             & 0 & 	0.0	         & 	0.0	     & $\mathrm{5s5p^3}$    & 	$^1\mathrm{P}^\mathrm{o}$	 & 1 & 	195068.2	 & 	194461.0	\\
                     &	$^3\mathrm{P}$	             & 1 & 	12543.7	     & 	12176.0	 &                      &	term	                     &   &	195068.2	 & 	194461.0	\\ 
                     &	$^3\mathrm{P}$	             & 2 & 	18176.0	     & 	17628.2	 &  $\mathrm{5s^25p5d}$ & 	$^3\mathrm{P}^\mathrm{o}$	 & 2 & 	198040.8	 & 	197579.4	\\ 
                     &	term		                 &   &  14279.0	     & 	13852.1	 &                      &	$^3\mathrm{P}^\mathrm{o}$	 & 0 & 	212559.6	 & 	212928.1	\\
$\mathrm{5s^25p^2}$  & 	$^1\mathrm{D}$	             & 2 & 	36082.9	     & 	35061.4	 &                      &	$^3\mathrm{P}^\mathrm{o}$	 & 1 & 	214569.6	 & 	214186.0	\\
                     &	term		                 &   &  36083.0	     & 	35061.4	 &                      &	term	                     &   &	205163.6	 & 	204820.3	\\
\hline
\end{tabular}
\end{minipage}
\end{table*}


\begin{table*}
\centering
\begin{minipage}{140mm}
\caption{\ion{Cs}{IV}: List of tabulated lines ordered in increasing wavelengths for allowed transitions that are given in Table \ref{tab:CsIV_allowed_transitions}.\label{tab:CsIV_list_lines} }
\begin{tabular}{@{}ccccccccccc@{}}
\hline
Wavelength (\AA) &  No. & & Wavelength (\AA) & No. & & Wavelength (\AA) & No. & & Wavelength (\AA) & No. \\
\hline
   469.387 &  25 & &   566.165 &  45 & &   630.026 &  38 & &   734.521 &   2 \\
   479.616 &  24 & &   566.605 &  17 & &   630.726 &  13 & &   735.403 &   2 \\
   480.913 &  23 & &   567.164 &  19 & &   631.129 &  41 & &   736.714 &   4 \\
   491.897 &  25 & &   567.651 &  18 & &   631.290 &  35 & &   744.648 &  31 \\
   499.645 &  25 & &   572.117 &  12 & &   633.696 &  10 & &   757.445 &  30 \\
   501.524 &  22 & &   572.349 &  12 & &   634.921 &  17 & &   759.570 &   3 \\
   503.143 &  24 & &   573.511 &  18 & &   637.179 &  13 & &   767.879 &  30 \\
\hline   
\end{tabular}
\end{minipage}
\end{table*}


\begin{table*}
\centering
\begin{minipage}{140mm}
\caption{\ion{Cs}{IV}: Allowed transitions (values of $A_{ki}$ and $f_{ik}$ are given with the notation $a\pm b\equiv a\times 10^{\pm b}$). \label{tab:CsIV_allowed_transitions}}
\begin{tabular}{@{}llcccccccr@{}}
\hline
No. & Transition & Mult. & $\lambda_{\rm{vac}}$ & $E_i$ & $E_k$ & $g_i$--$g_k$ & $A_{ki}$ & $f_{ik}$ & $\log gf$   \\
       & Array &  & (\AA) & (cm$^{-1}$) & (cm$^{-1}$) &   &  ($10^8$ s$^{-1}$) &  &  \\
\hline
1. & 5s$^2$5p$^4$-- & $^3$P -- $^3$P$^\circ$  & 886.789 & 5383.9 &  118150.2 & 9--9 &   5.230+00 & 6.097$-$02 & $-0.261$  \\
    &  $\;$ 5s5p$^5$ &   &     &     &     &    &     &   &    \\
    &  &    & 874.832 & 0.0 & 114307.7 & 5--5 & 3.729+00 & 4.212$-$02 & $-0.677$ \\
    &  &    & 824.793 & 0.0 & 121242.5 & 5--3 & 3.416+00 & 2.068$-$02 & $-0.985$ \\
    &  &    & 896.916 & 9749.4 & 121242.5 & 1--3 & 1.366+00 & 4.908$-$02 & $-1.309$ \\
    &  &    & 986.139 & 12902.0 & 114307.7 & 3--5 & 1.314+00 & 3.166$-$02 & $-1.022$ \\
\hline
\end{tabular}
\end{minipage}
\end{table*}

\begin{table*}
\centering
\begin{minipage}{140mm}
\contcaption{}
\begin{tabular}{@{}llcccccccr@{}}
\hline
No. & Transition & Mult. & $\lambda_{\rm{vac}}$ & $E_i$ & $E_k$ & $g_i$--$g_k$ & $A_{ki}$ & $f_{ik}$ & $\log gf$   \\
       & Array &  & (\AA) & (cm$^{-1}$) & (cm$^{-1}$) &   &  ($10^8$ s$^{-1}$) &  &  \\
\hline
10. & 5s$^2$5p$^4$-- & $^3$P -- $^3\rm{S}^\circ$  & 616.637 & 5383.9 & 167553.8 & 9--3 &  4.224+00 & 8.050$-$03 & $-1.140$ \\
      &  $\;$ 5s$^2$5p$^3$6s &  &  &  &   &  &  &   &   \\
      &  &   & 596.823 & 0.0 & 167553.8 & 5--3 & 3.083+00 & 9.896$-$03 & $-1.306$ \\
      &  &   & 633.696 & 9749.4 & 167553.8 & 1--3 & 1.215+00 & 2.205$-$02 & $-1.657$ \\
      &  &   & 646.614 & 12902.0 & 167553.8 & 3--3 & 9.564$-$02 & 6.040$-$04 & $-2.742$ \\
    &  &    &               &               &                 &        &  &  &   \\
\hline
\end{tabular}
\end{minipage}
\end{table*}

\begin{table*}
\centering
\begin{minipage}{140mm}
\contcaption{}
\begin{tabular}{@{}llcccccccr@{}}
\hline
No. & Transition & Mult. & $\lambda_{\rm{vac}}$ & $E_i$ & $E_k$ & $g_i$--$g_k$ & $A_{ki}$ & $f_{ik}$ & $\log gf$   \\
       & Array &  & (\AA) & (cm$^{-1}$) & (cm$^{-1}$) &   & ($10^8$ s$^{-1}$) & & \\
\hline
18. & 5s$^2$5p$^4$-- & $^3$P -- $^3\rm{P}^\circ$  & 549.259 & 5383.9 & 187447.4 & 9--9 &   2.579+02 & 1.166+00 & 1.021 \\
       &  $\;$ 5s$^2$5p$^3$5d &  &  &  &   &  &  &   &   \\
       &  &  & 533.998 & 0.0 & 187266.6 & 5--5 & 2.286+02 & 1.005+00 & 0.701 \\
       &  &  & 537.883 & 0.0 & 185913.9 & 5--3 & 1.019+02 & 2.580$-$01 & 0.111 \\
       &  &  & 567.651 & 9749.4 & 185913.9 & 1--3 & 9.469+01 & 1.336+00 & 0.126 \\
       &  &  & 573.511 & 12902.0 & 187266.6 & 3--5 & 4.621+00 & 3.933$-$02 & $-0.928$ \\
\hline
\end{tabular}
\end{minipage}
\end{table*}

\begin{table*}
\centering
\begin{minipage}{140mm}
\contcaption{}
\begin{tabular}{@{}llcccccccr@{}}
\hline
No. & Transition & Mult. & $\lambda_{\rm{vac}}$ & $E_i$ & $E_k$ & $g_i$--$g_k$ & $A_{ki}$ & $f_{ik}$ & $\log gf$   \\
       & Array &  & (\AA) & (cm$^{-1}$) & (cm$^{-1}$) &   &  ($10^8$ s$^{-1}$) &  &  \\
\hline
28. & 5s$^2$5p$^4$-- & $^1$D -- $^1\rm{P}^\circ$  & 828.851 & 20754.0 & 141402.9 & 5--3 & 5.709+00 & 3.662$-$02 & $-0.737$ \\
       &  $\;$ 5s$^2$5p$^3$5d &  &  &  &   &  &  &   &   \\
       &  &   &               &               &                 &        &  &  &   \\
29. & 5s$^2$5p$^4$-- & $^1$D -- $^3\rm{D}^\circ$  & 798.164 & 20754.0 & 146041.6 & 5--15 & 2.768$-$02 & 7.930$-$04 & $-2.402$ \\
       &  $\;$ 5s$^2$5p$^3$5d &  &  &  &   &  &  &   &   \\
       &  &  & 823.635 & 20754.0 & 142167.0 & 5--5 & 6.335-02 & 6.438$-$04 & $-2.492$ \\
\hline
\end{tabular}
\end{minipage}
\end{table*}

\begin{table*}
\centering
\begin{minipage}{140mm}
\contcaption{}
\begin{tabular}{@{}llcccccccr@{}}
\hline
No. & Transition & Mult. & $\lambda_{\rm{vac}}$ & $E_i$ & $E_k$ & $g_i$--$g_k$ & $A_{ki}$ & $f_{ik}$ & $\log gf$   \\
       & Array &  & (\AA) & (cm$^{-1}$) & (cm$^{-1}$) &   &  ($10^8$ s$^{-1}$) &  &  \\
\hline
42. & 5s$^2$5p$^4$-- & $^1$D -- $^3\rm{P}^\circ$  & 599.904 & 20754.0 & 187447.4 & 5--9 & 2.509+01 &  2.372$-$01 & $0.074$ \\
       &  $\;$ 5s$^2$5p$^3$5d &  &  &  &   &  &  &   &   \\
       &  &  & 600.555 & 20754.0 & 187266.6 & 5--5 & 7.208$-$01 & 4.038$-$03 & $-1.695$ \\
       &  &  & 605.474 & 20754.0 & 185913.9 & 5--3 & 7.204+01 & 2.310$-$01 & $0.063$ \\
       &  &   &               &               &                 &        &  &  &   \\
43. & 5s$^2$5p$^4$-- & $^1$D -- $^1\rm{D}^\circ$  & 593.599 & 20754.0 & 189217.8 & 5--5 & 3.339+01 & 1.773$-$01 & $-0.052$ \\
\hline
\end{tabular}
\end{minipage}
\end{table*}

\begin{table*}
\centering
\begin{minipage}{140mm}
\contcaption{}
\begin{tabular}{@{}llcccccccr@{}}
\hline
No. & Transition & Mult. & $\lambda_{\rm{vac}}$ & $E_i$ & $E_k$ & $g_i$--$g_k$ & $A_{ki}$ & $f_{ik}$ & $\log gf$   \\
       & Array &  & (\AA) & (cm$^{-1}$) & (cm$^{-1}$) &   &  ($10^8$ s$^{-1}$) &  &  \\
\hline
59. & 5s$^2$5p$^4$-- & $^1$S -- $^3$D$^\circ$  & 713.212 & 43279.4 & 183490.2 & 1--15 &  5.199$-$02 & 5.614$-$03 & $-2.251$  \\
       & $\;$ 5s$^2$5p$^3$5d &  &  &  &   &  &  &   &   \\ 
       &  &  & 684.405 & 43279.4 & 189391.7 & 1--3 & 2.942$-$01 & 5.850$-$03 & $-2.233$ \\
       &  &   &               &               &                 &        &  &  &   \\
60. & 5s$^2$5p$^4$-- & $^1$S -- $^3$P$^\circ$  & 693.635 & 43279.4 & 187447.4 & 1--9 &  1.070$-$01 & 6.741$-$03 & $-2.171$ \\
       &  $\;$ 5s$^2$5p$^3$5d &  &  &  &   &  &  &   &   \\
\hline
\end{tabular}
\end{minipage}
\end{table*}


\begin{table*}
\centering
\begin{minipage}{140mm}
\caption{\ion{Cs}{V}: List of tabulated lines ordered in increasing wavelengths for allowed transitions that are given in Table \ref{tab:CsV_allowed_transitions}.\label{tab:CsV_list_lines}}
\begin{tabular}{@{}ccccccccccc@{}}
\hline
Wavelength (\AA) &  No. & & Wavelength (\AA) & No. & & Wavelength (\AA) & No. & & Wavelength (\AA) & No. \\
\hline
   421.337 &  18 & & 522.743 &  30 & & 591.050 &  46 & &  700.154 &  23 \\
   425.183 &  18 & & 522.912 &   8 & & 591.567 &  27 & &  714.360 &  43 \\
   446.427 &  17 & & 523.279 &  31 & & 592.379 &  51 & &  714.494 &   2 \\
   446.749 &  14 & & 524.354 &  33 & & 594.628 &  49 & &  714.519 &  23 \\
   449.918 &  37 & & 525.967 &  34 & & 595.691 &  28 & &  714.828 &  22 \\
   450.585 &  15 & & 526.034 &  36 & & 597.921 &  26 & &  726.708 &  43 \\
   454.307 &  37 & & 526.338 &   8 & & 599.520 &   6 & &  727.117 &  23 \\
\hline
\end{tabular}
\end{minipage}
\end{table*}


\begin{table*}
\centering
\begin{minipage}{140mm}
\caption{\ion{Cs}{V}: Allowed transitions (values of $A_{ki}$ and $f_{ik}$ are given with the notation $a\pm b\equiv a\times 10^{\pm b}$). \label{tab:CsV_allowed_transitions}}
\begin{tabular}{@{}llcccccccr@{}}
\hline
No. & Transition & Mult. & $\lambda_{\rm{vac}}$ & $E_i$ & $E_k$ & $g_i$--$g_k$ & $A_{ki}$ & $f_{ik}$ & $\log gf$   \\
       & Array &  & (\AA) & (cm$^{-1}$) & (cm$^{-1}$) &   &  ($10^8$ s$^{-1}$) &  &  \\
\hline
1. & 5s$^2$5p$^3$-- & $^4$S$^\circ$--$^4$P  & 840.397 & 0.0 & 118991.4 & 4--12 & 4.292+00 & 1.361$-$01 & $-0.264$ \\
       &  $\;$ 5s5p$^4$ &  &  &  &   &  &  &   &   \\ 
       &  &  & 877.950 & 0.0 & 113901.7 & 4--6 & 3.382+00 & 5.855$-$02 & $-0.630$ \\
       &  &  & 811.738 & 0.0 & 123192.4 & 4--4 & 5.165+00 & 5.075$-$02 & $-0.693$ \\
       &  &  & 794.544 & 0.0 & 125858.3 & 4--2 & 5.766+00 & 2.737$-$02 & $-0.961$ \\
       &  &   &               &               &                 &        &  &  &   \\
\hline
\end{tabular}
\end{minipage}
\end{table*}

\begin{table*}
\centering
\begin{minipage}{140mm}
\contcaption{}
\begin{tabular}{@{}llcccccccr@{}}
\hline
No. & Transition & Mult. & $\lambda_{\rm{vac}}$ & $E_i$ & $E_k$ & $g_i$--$g_k$ & $A_{ki}$ & $f_{ik}$ & $\log gf$   \\
       & Array &  & (\AA) & (cm$^{-1}$) & (cm$^{-1}$) &   &  ($10^8$ s$^{-1}$) &  &  \\
\hline
13. & 5s$^2$5p$^3$-- & $^4$S$^\circ$--$^2$P  & 462.721 & 0.0 & 216113.1 & 4--6 & 2.954+00 & 1.398$-$02 & $-1.252$ \\
       &  $\;$ 5s$^2$5p$^2$6s &  &  &  &   &  &  &   &   \\ 
       &  &  & 466.934 & 0.0 & 214163.1 & 4--4 & 1.393$-$02 & 4.482$-$05 & $-3.746$ \\
       &  &  & 454.518 & 0.0 & 220013.2 & 4--2 & 9.319+00 & 1.419$-$02 & $-1.246$ \\
       &  &               &               &                 &        &  &  &   \\ 
14. & 5s$^2$5p$^3$-- & $^4$S$^\circ$--$^2$D  & 462.564 & 0.0 & 216186.4 & 4--10 & 4.321+00 & 3.466$-$02 & $-0.858$ \\
\hline
\end{tabular}
\end{minipage}
\end{table*}

\begin{table*}
\centering
\begin{minipage}{140mm}
\contcaption{}
\begin{tabular}{@{}llcccccccr@{}}
\hline
No. & Transition & Mult. & $\lambda_{\rm{vac}}$ & $E_i$ & $E_k$ & $g_i$--$g_k$ & $A_{ki}$ & $f_{ik}$ & $\log gf$   \\
       & Array &  & (\AA) & (cm$^{-1}$) & (cm$^{-1}$) &   &  ($10^8$ s$^{-1}$) &  &  \\
\hline
24. & 5s$^2$5p$^3$-- & $^2$D$^\circ$--$^4$D & 660.403 & 18255.1 & 169677.8 & 10--20 & 2.443+00 & 3.271$-$02 & $-0.485$ \\
       &  $\;$ 5s$^2$5p$^2$5d &  &  &  &   &  &  &   &   \\ 
       &  &  & 647.331 & 15077.4 & 169557.9 & 4--2 & 3.354+00 & 1.108$-$02 & $-1.354$ \\
       &  &  & 659.097 & 15077.4 & 166800.2 & 4--6 & 3.225$-$01 & 3.080$-$03 & $-1.909$ \\
       &  &  & 682.936 & 20373.5 & 166800.2 & 6--6 & 9.962$-$01 & 6.810$-$03 & $-1.389$ \\
       &  &  & 637.514 & 15077.4 & 171936.7 & 4--4 & 2.002+00 & 1.264$-$02 & $-1.296$ \\
\hline
\end{tabular}
\end{minipage}
\end{table*}

\begin{table*}
\centering
\begin{minipage}{140mm}
\contcaption{}
\begin{tabular}{@{}llcccccccr@{}}
\hline
No. & Transition & Mult. & $\lambda_{\rm{vac}}$ & $E_i$ & $E_k$ & $g_i$--$g_k$ & $A_{ki}$ & $f_{ik}$ & $\log gf$   \\
       & Array &  & (\AA) & (cm$^{-1}$) & (cm$^{-1}$) &   &  ($10^8$ s$^{-1}$) &  &  \\
\hline
33. & 5s$^2$5p$^3$-- & $^2$D$^\circ$--$^2$D & 505.226 & 18255.1 & 216186.4 & 10--10 & 1.130+02 & 4.323$-$01 & $0.636$ \\
       &  $\;$ 5s$^2$5p$^2$5d &  &  &  &   &  &  &   &   \\
       &  &  & 510.186 & 15077.4 & 211084.4 & 4--6 & 1.318+02 & 7.717$-$01 & $0.490$ \\
       &  &  & 524.354 & 20373.5 & 211084.4 & 6--6 & 4.408+01 & 1.820$-$01 & $0.038$ \\
       &  &  & 479.014 & 15077.4 & 223838.4 & 4--4 & 1.444$-$01 & 4.732$-$04 & $-2.723$ \\
       &  &  & 491.483 & 20373.5 & 223838.4 & 6--4 & 5.217+00 & 1.200$-$02 & $-1.143$ \\
\hline
\end{tabular}
\end{minipage}
\end{table*}

\begin{table*}
\centering
\begin{minipage}{140mm}
\contcaption{}
\begin{tabular}{@{}llcccccccr@{}}
\hline
No. & Transition & Mult. & $\lambda_{\rm{vac}}$ & $E_i$ & $E_k$ & $g_i$--$g_k$ & $A_{ki}$ & $f_{ik}$ & $\log gf$   \\
       & Array &  & (\AA) & (cm$^{-1}$) & (cm$^{-1}$) &   &  ($10^8$ s$^{-1}$) &  &  \\
\hline
43. & 5s$^2$5p$^3$-- & $^2$P$^\circ$--$^4$D & 764.263 & 38832.8 & 169677.8 & 6--20 & 4.110$-$01 & 1.263$-$02 & $-1.120$ \\
       &  $\;$ 5s$^2$5p$^2$5d &  &  &  &   &  &  &   &   \\
       &  &  & 726.708 & 31951.1 & 169557.9 & 2--2 & 2.472+00 & 2.064$-$02 & $-1.384$ \\
       &  &  & 785.643 & 42273.7 & 169557.9 & 4--2 & 1.066+00 & 5.252$-$03 & $-1.678$ \\
       &  &  & 803.042 & 42273.7 & 166800.2 & 4--6 & 3.655$-$02 & 5.168$-$04 & $-2.685$ \\
       &  &  & 714.360 & 31951.1 & 171936.7 & 2--4 & 1.991$-$01 & 3.160$-$03 & $-2.199$ \\
\hline
\end{tabular}
\end{minipage}
\end{table*}

\begin{table*}
\centering
\begin{minipage}{140mm}
\contcaption{}
\begin{tabular}{@{}llcccccccr@{}}
\hline
No. & Transition & Mult. & $\lambda_{\rm{vac}}$ & $E_i$ & $E_k$ & $g_i$--$g_k$ & $A_{ki}$ & $f_{ik}$ & $\log gf$   \\
       & Array &  & (\AA) & (cm$^{-1}$) & (cm$^{-1}$) &   &  ($10^8$ s$^{-1}$) &  &  \\
\hline
52. & 5s$^2$5p$^3$-- & $^2$P$^\circ$--$^2$P & 563.304 & 38832.8 & 216356.9 & 6--6 & 1.358+02 & 6.209$-$01 & $0.571$ \\
       &  $\;$ 5s$^2$5p$^2$5d &  &  &  &   &  &  &   &   \\
       &  &  & 577.194 & 31951.1 & 205203.2 & 2--2 & 6.991+01 & 3.391$-$01 & $-0.169$ \\
       &  &  & 613.762 & 42273.7 & 205203.2 & 4--2 & 3.168+00 & 8.710$-$03 & $-1.458$ \\
       &  &  & 526.364 & 31951.1 & 221933.8 & 2--4 & 5.556+01 & 4.422$-$01 & $-0.053$ \\
       &  &  & 556.606 & 42273.7 & 221933.8 & 4--4 & 1.230+02 & 5.480$-$01 & 0.341 \\
\hline
\end{tabular}
\end{minipage}
\end{table*}


\begin{table*}
\centering
\begin{minipage}{140mm}
\caption{\ion{Cs}{VI}: List of tabulated lines ordered in increasing wavelengths for allowed transitions that are given in Table \ref{tab:CsVI_allowed_transitions}.\label{tab:CsVI_list_lines}}
\begin{tabular}{@{}ccccccccccc@{}}
\hline
Wavelength (\AA) &  No. & & Wavelength (\AA) & No. & & Wavelength (\AA) & No. & & Wavelength (\AA) & No. \\
\hline
   378.463 &  14 & &    500.504 &  10 & &    572.299 &  23 & &    729.141 &   3 \\
   396.746 &  14 & &    504.097 &  10 & &    589.477 &   6 & &    732.225 &   3 \\
   401.968 &  13 & &    506.024 &   9 & &    589.695 &   6 & &    748.118 &  18 \\
   405.517 &  14 & &    507.734 &  26 & &    607.027 &  24 & &    758.917 &   2 \\
   410.308 &  13 & &    508.756 &   8 & &    609.284 &   6 & &    811.458 &  31 \\
   410.975 &  13 & &    514.090 &  10 & &    611.736 &   5 & &    827.038 &   2 \\
   431.007 &  12 & &    514.242 &   7 & &    615.316 &  22 & &    830.464 &   2 \\
   431.885 &  13 & &    520.381 &   9 & &    618.140 &  33 & &    835.321 &  17 \\
   434.712 &  13 & &    522.297 &  25 & &    627.354 &  21 & &    836.185 &   2 \\
   436.366 &  28 & &    522.715 &  36 & &    632.843 &   5 & &    839.371 &  17 \\
   442.300 &  13 & &    532.985 &  10 & &    638.756 &   4 & &    866.092 &   2 \\
   442.692 &  27 & &    539.364 &   8 & &    648.516 &   3 & &    876.128 &   2 \\
   454.879 &  12 & &    548.592 &   7 & &    656.993 &  20 & &    971.049 &  16 \\
   466.447 &  12 & &    548.937 &  10 & &    661.804 &   4 & &    982.436 &  30 \\
   466.884 &   8 & &    552.665 &  24 & &    678.480 &  34 & &   1020.117 &  16 \\
   472.107 &  37 & &    555.706 &   8 & &    681.692 &  20 & &   1034.069 &  16 \\
   478.709 &  11 & &    556.779 &  35 & &    701.263 &   3 & &   1055.938 &   1 \\
   479.254 &  27 & &    558.271 &  22 & &    703.974 &  32 & &   1120.444 &   1 \\
   490.613 &  10 & &    564.700 &  24 & &    704.116 &   3 & &   1260.137 &  29 \\
   495.025 &   8 & &    565.506 &   7 & &    711.319 &  19 & &   1392.425 &  15 \\
   498.127 &   8 & &    569.329 &   5 & &    711.953 &   3 & &            &     \\
\hline
\end{tabular}
\end{minipage}
\end{table*}


\begin{table*}
\centering
\begin{minipage}{140mm}
\caption{\ion{Cs}{VI}: Allowed transitions (values of $A_{ki}$ and $f_{ik}$ are given with the notation $a\pm b\equiv a\times 10^{\pm b}$). \label{tab:CsVI_allowed_transitions}}
\begin{tabular}{@{}llcccccccr@{}}
\hline
No. & Transition & Mult. & $\lambda_{\rm{vac}}$ & $E_i$ & $E_k$ & $g_i$--$g_k$ & $A_{ki}$ & $f_{ik}$ & $\log gf$   \\
       & Array &  & (\AA) & (cm$^{-1}$) & (cm$^{-1}$) &   &  ($10^8$ s$^{-1}$) &  &  \\
\hline
1. & 5s$^2$5p$^2$-- & $^3$P -- $^5\rm{S}^\circ$  & 1074.963 & 13852.1 & 106878.6 & 9--5 &  5.609$-$01 & 5.383$-$03 & $-1.315$ \\
       &  $\;$ 5s5p$^3$ &  &  &  &   &  &  &   &   \\ 
       &  &  & 1055.938 & 12176.0 & 106878.6 & 3--5 & 2.878$-$01 & 7.980$-$03 & $-1.621$ \\
       &  &  & 1120.444 & 17628.2 & 106878.6 & 5--5 & 2.544$-$01 & 4.784$-$03 & $-1.621$ \\
       &  &   &               &               &                 &        &  &  &   \\ 
2. & 5s$^2$5p$^2$-- & $^3$P -- $^3\rm{D}^\circ$  & 824.508 & 13852.1 & 135136.5 & 9--15 &  4.920+00 & 8.321$-$02 & $-0.126$ \\
\hline
\end{tabular}
\end{minipage}
\end{table*}

\begin{table*}
\centering
\begin{minipage}{140mm}
\contcaption{}
\begin{tabular}{@{}llcccccccr@{}}
\hline
No. & Transition & Mult. & $\lambda_{\rm{vac}}$ & $E_i$ & $E_k$ & $g_i$--$g_k$ & $A_{ki}$ & $f_{ik}$ & $\log gf$   \\
       & Array &  & (\AA) & (cm$^{-1}$) & (cm$^{-1}$) &   &  ($10^8$ s$^{-1}$) &  &  \\
\hline
10. & 5s$^2$5p$^2$-- & $^3$P -- $^3\rm{D}^\circ$  & 507.330 & 13852.1 & 210962.4 & 9--15 & 2.298+02 & 1.462+00 & $1.119$ \\
       &  $\;$ 5s$^2$5p5d &  &  &  &   &  &  &   &   \\
       &  &  & 500.504 & 0.0 & 199798.6 & 1--3 & 2.379+02 & 2.631+00 & $0.420$ \\
       &  &  & 532.985 & 12176.0 & 199798.6 & 3--3 & 5.529+01 & 2.318$-$01 & $-0.158$ \\
       &  &  & 548.937 & 17628.2 & 199798.6 & 5--3 & 4.886+00 & 1.305$-$02 & $-1.185$ \\
       &  &  & 514.090 & 17628.2 & 212146.6 & 5--7 & 2.754+02 & 1.512+00 & $0.879$ \\
\hline
\end{tabular}
\end{minipage}
\end{table*}

\begin{table*}
\centering
\begin{minipage}{140mm}
\contcaption{}
\begin{tabular}{@{}llcccccccr@{}}
\hline
No. & Transition & Mult. & $\lambda_{\rm{vac}}$ & $E_i$ & $E_k$ & $g_i$--$g_k$ & $A_{ki}$ & $f_{ik}$ & $\log gf$   \\
       & Array &  & (\AA) & (cm$^{-1}$) & (cm$^{-1}$) &   &  ($10^8$ s$^{-1}$) &  &  \\
\hline
22. & 5s$^2$5p$^2$-- & $^1$D -- $^3\rm{P}^\circ$  &  589.071 & 35061.4 & 204820.3 & 5--9 & 2.753+01 & 2.596$-$01 & $0.113$\\ 
       &  $\;$ 5s$^2$5p5d &  &  &  &   &  &  &   &   \\
       &  &  & 615.316 & 35061.4 & 197579.4 & 5--5 & 2.202+01 & 1.259$-$01 & $-0.201$ \\
       &  &  & 558.271 & 35061.4 & 214185.9 & 5--3 & 4.788+01 & 1.352$-$01 & $-0.170$ \\
       &  &   &               &               &                 &        &  &  &   \\ 
23. & 5s$^2$5p$^2$-- & $^1$D -- $^1\rm{D}^\circ$  & 572.299 & 35061.4 & 209795.3 & 5--5 &6.031+01 & 3.013$-$01 & $0.178$ \\
\hline
\end{tabular}
\end{minipage}
\end{table*}


\bsp	
\label{lastpage}
\end{document}